# Reversible room-temperature magnetocaloric effect with large temperature span in antiperovskite compounds $Ga_{1-x}CMn_{3+x}$ ($x$=0, 0.06, 0.07, and 0.08)

B. S. Wang, P. Tong,[a)] Y. P. Sun,[b)] X. B. Zhu, X. Luo, G. Li, W. H. Song, Z. R. Yang and J. M. Dai

*Key Laboratory of Materials Physics, Institute of Solid State Physics, and High Magnetic Field Laboratory, Chinese Academy of Sciences, Hefei 230031, People's Republic of China*

**Abstract**

The magnetic and magnetocaloric properties of $Ga_{1-x}CMn_{3+x}$ have been investigated. Reversible magnetocaloric effect (MCE) occurs near the Curie temperature $T_C$. With increasing $x$, we find that the magnetic entropy change $-\Delta S_M$ decreases while $T_C$ and magnetization increase; Meanwhile, the temperature span of $-\Delta S_M$ vs. $T$ plot becomes well broadened. Due to the competition between the broadening temperature span and decreasing $-\Delta S_M$, the relative cooling power (RCP) increases initially, and then decreases with increasing $x$ further. The largest RCP (2.1 J/cm$^3$, in a magnetic field of 45 kOe) observed at $x$=0.07 ($T_C$ = 296.5 K) is comparable to the contemporary magnetic refrigerant materials. Considering the reversible MCE, inexpensive and innoxious raw materials, our result suggests that $Ga_{1-x}CMn_{3+x}$ can be a promising candidate for magnetic refrigeration around room temperature.

[a)] Electronic mail: tongpeng@issp.ac.cn
[b)] Electronic mail: ypsun@issp.ac.cn

# 1. Introduction

The magnetocaloric effect (MCE) is a process in which an isothermal variation of the magnetic entropy change takes place with an adiabatic temperature change of the system on the application of a magnetic field. MCE is at the heart of magnetic refrigeration and expected to be competitive with the traditional gas compression technology because of its higher energy-efficiency and less impact on environment.[1-3] Large or giant MCE has been observed on the base of both first-order and second-order magnetic phase transitions.[2] However, the practical application of first-order magnetic phase transition is restricted because the relevant hysteretic properties may deteriorate magnetocaloric parameters during the cyclic process of magnetization-demagnetization.[2] Moreover, most of the promising magnetic refrigerant materials are made of expensive rare earth metals (e.g., Gd) or poisonous elements (e.g., As). Therefore, it is of great importance to search for materials with reversible MCE, as well as low-cost and innoxious raw materials.

Recently, the antiperovskite compounds have attracted great attention due to their interesting properties, such as superconductivity,[4,5] non-Fermi liquid behavior,[6] strong electron correlations,[7] giant magnetoresistance (GMR),[8,9] MCE,[10-13] and giant negative thermal expansion (NTE).[14,15] It is known that $GaCMn_3$, a typical antiperovskite-structured compound, exhibits three magnetic transitions: first-order transition occurs from antiferromagnetic (AFM) ground state to an intermediate canted ferromagnetic (CF) state with an abrupt decrease in volume at 160 K; second-order transition appears from CF to ferromagnetic (FM) phase at 164 K, which will be suppressed under high magnetic fields; and the other second-order transition happens from FM to paramagnetic (PM) state at 246 K.[8] MCE was observed around the first-order AFM-CM transition with a large entropy change $-\Delta S_M$ (15 J/kg K under 20 kOe), which has a plateau-like temperature dependence.[10,11] This MCE is sensitive to the chemical composition. For instance, in $GaCMn_{3-x}Co_x$, the substitution of Co for Mn sites decreases the AFM-FM transition temperature without a significant loss of $-\Delta S_M$, but broadens the plateau-like temperature dependence.[12] In carbon-deficient sample $GaC_{0.78}Mn_3$, the AFM state observed in stoichiometric sample is collapsed. Instead, the ground state is FM with a transition temperature of 295 K, at which the maximum value of $-\Delta S_M$ is 3.7 J/kg K under 50 kOe.[13] So far, the magnetic entropy change with respect to the second-order FM-PM transition has not been reported in detail, which is possibly due to its small magnetic entropy change.

In this paper, we report the preparation, structural and magnetic properties of $Ga_{1-x}CMn_{3+x}$ ($x$=0, 0.06, 0.07, and 0.08). The second-order FM-PM transition temperature $T_C$ is found to increase from 250 K for $x$=0 to 323.5 K for $x$=0.08, which can be attributed to the enhancement of FM exchange interaction due to the reduction of Mn-Mn atomic distance. Unlike the first-order AFM-CM transition, no detectable magnetic or thermal hysteresis can be found round $T_C$. As $x$ increases, the maximum entropy change decreases. However, the peak in the temperature dependence of the entropy change is broadened, leading to a general increase of the relative cooling power (RCP). The reversible behavior of magnetization, enhanced RCP with large temperature span, and low-cost raw materials suggest potential application of $Ga_{1-x}CMn_{3+x}$ with $x$=0.06, 0.07 and 0.08 in the field of magnetic refrigeration near room temperature.

**2. Experimental details**

Polycrystalline samples of $Ga_{1-x}CMn_{3+x}$ were prepared from pieces of Ga (4N) and powders of Graphite (3N) and Mn (4N). The starting materials were mixed in the desired proportions, sealed in evacuated quartz tubes, and then heated at 1073K-1123K for eight days. After the tubes were quenched to room temperature, the products were pulverized, mixed, and annealed again under the same condition in order to obtain the homogeneous samples. X-ray diffraction (XRD) measurements on powder samples were performed by using Cu Kα radiation to identify the phase purity and the crystal structure. Magnetizations were measured as a function of temperature and magnetic field on a Quantum Design superconducting quantum interference device (SQUID) magnetometer (1.8 K ≤ $T$ ≤ 400 K, 0 ≤ $H$ ≤ 50 kOe).

**3. Results and discussion**

Figure 1 shows the room-temperature powder XRD patterns of $Ga_{1-x}CMn_{3+x}$ ($x$=0.0, 0.06, 0.07, 0.08). All samples have a single phase with a standard cubic antiperovskite structure, except for a small amount of unreacted graphite. As shown in the inset of Fig. 1, the (220) peak shifts toward higher angles as $x$ increases. Accordingly, the refined lattice constant $a$ decreases with increasing the doping density $x$ (Fig. 3).

Figure 2 shows the temperature dependence of magnetization of $Ga_{1-x}CMn_{3+x}$ at a magnetic field of 100 Oe under both the field cooling cool (FCC) and field cooling warming (FCW) modes.

It is apparent that there is no observable hysteresis during cooling and warming cycles, which indicates the existence of a second-order transition. The Curie temperatures $T_C$ determined from the derivative of $M(T)$ curves are found to be 250K, 281.5K, 296.5K, and 323.5K for $x$=0, 0.06, 0.07, and 0.08, respectively. We also found that the magnetization below $T_C$ also increases with increasing $x$. As displayed in Fig. 3, the enhancement of $T_C$ can be attributed to the lattice contraction, which leads to an increased exchange interaction among Mn atoms.[9]

As shown in Fig. 4, the isotherm magnetization curves $M(H)$ of $Ga_{1-x}CMn_{3+x}$ were measured at 214 K (which locates well in the FM state) under the magnetic fields up to 45kOe. The magnetization increases sharply at low fields, and then tends to be saturated at above 10 kOe. The extrapolation of the high field $M(H)$ curve to $M$-axis yields a saturated magnetization which increases with increasing $x$ (to see the inset of Fig. 4), in agreement with the evolution of $T_C$. The $M(H)$ curves around $T_C$ are plotted in Fig. 5 (*a*), (*b*), (*c*) and (*d*) for the samples with $x$=0, 0.06, 0.07 and 0.08, respectively. All curves are reversible during the increasing/decreasing field cycling without hysteresis. Such a non-hysteresis behavior of $M(H)$ and $M(T)$ is significant for the practical application of the magnetic refrigeration.[2] The Arrott plots derived from $M(H)$ within a broad temperature range around $T_C$ are displayed in Fig. 6. For all samples, it is evident that the slope of $H/M$ vs $M^2$ curves at high fields is positive at all temperatures measured, which confirms the second-order magnetic transition according to the Banerjee criterion.[16] In the Arrott plots, the curve $H/M$ vs $M^2$ measured at $T_C$ crosses over the origin (see the red lines in Fig. 6).[17] In this way, the determined $T_C$ for each sample is basically identical to that obtained from the derivative of the $M(T)$ curve (Fig. 2).

According to the classical thermo-dynamical theory, $-\Delta S_M$ induced by the variation of a magnetic field from 0 to $H$ is given by[18]

$$\Delta S_M(T,H) = \int_0^H \left(\frac{\partial M}{\partial T}\right)_H dH . \tag{1}$$

For the magnetization measured at a small discrete field and temperature intervals, $-\Delta S_M$ can be approximated as[18]

$$\left|\Delta S_M\left(\frac{T_i + T_{i+1}}{2}\right)\right| = \sum \left[\frac{(M_i - M_{i+1})_{H_i}}{T_{i+1} - T_i}\right]\Delta H_i , \tag{2}$$

where $M_i$ and $M_{i+1}$ are the experimental values of the magnetization at $T_i$ and $T_{i+1}$ under the same magnetic field, respectively. From Eq. (2), $-\Delta S_M$ associated with the magnetic field variation can be calculated from the isothermal $M(H)$ curves at various temperatures. As shown in Fig. 7, $-\Delta S_M$ is plotted as a function of temperature for all samples at $\Delta H =$ 20kOe and 45kOe. As expected from Eq. (2), the peak of $-\Delta S$ appears around $T_C$ where the variation of moment with temperature is the sharpest. The maximum $-\Delta S_M$ value ($-\Delta S_M^{max}$) under $\Delta H =$ 45 kOe is about 4.19, 2.65, 2.39 and 1.81 J/kg K for $x =$ 0, 0.06, 0.07, and 0.08, respectively. At each $\Delta H$, $-\Delta S_M^{max}$ is reduced as $x$ increases. Interestingly, we find that the width of the peak, namely the temperature span is broadened with increasing $x$. For $x =$ 0.08, under each $\Delta H$ the temperature dependence of $-\Delta S_M$ has a table-like shape with a span of 160 K which covers the room temperature. Such a table-like shape in the temperature dependence of $-\Delta S_M$ is quite useful to the application of magnetic refrigerants. Further, $-\Delta S_M^{max}$ at $\Delta H=$45 kOe is obviously enhanced compared with that at $\Delta H=$20 kOe.

In the framework of the mean field theory, the relation between maximum magnetic entropy change and the magnetic field near $T_C$ is described as[19]

$$\Delta S_M^{max} \approx -1.07 qR \left( \frac{g\mu_B JH}{kT_C} \right)^{2/3} \text{ with } M_S = g\mu_B J, \tag{3}$$

where $q$ is the number of magnetic ions, $R$ is the gas constant, and $g$ is the Lande factor. The $H^{2/3}$ dependence of $-\Delta S_M^{max}$ is depicted in Fig. 8. It can be seen that $-\Delta S_M^{max}$ is linearly dependent on $H^{2/3}$ for all $Ga_{1-x}CMn_{3+x}$ samples, indicating the second-order character of magnetic transition under the framework of the mean field theory.[20] This result is in good agreement with those deduced from Fig. 2, 4, 5 and 6.

RCP, a measure of how much heat can be transferred between the cold and hot sinks in an ideal refrigerant cycle,[2] is one of the most important parameters for selecting potential substances for magnetic refrigerants. Generally, the RCP is defined as the product of the maximum magnetic entropy change $-\Delta S_M^{max}$ and the full width at half maximum $\delta T_{FWHM}$,[21]

$$RCP = -\Delta S_M^{max} \delta T_{FWHM}. \tag{4}$$

For $Ga_{1-x}CMn_{3+x}$, $\delta T_{FWHM}$ increases along with increasing $x$, or the external field $H$, as shown in Fig. 9 (*a*). At 45 kOe, $\delta T_{FWHM}$ reaches 110 K, 126 K, and 164 K for $x$ = 0.06, 0.07 and 0.08, respectively. These values are much larger than that of Gd (~60 K) under 50 kOe.[22] As indicated in Fig. 9 (*b*), on one hand, the RCP increases with increasing $H$ for each composition owing to the increased $-\Delta S_M^{max}$ by increasing $H$ (see Fig. 8). On the other hand, RCP does not increase monotonously with $x$. At 45 kOe, for example, RCP increases with increasing $x$ from 0 to 0.07, but decreases with increasing $x$ further, as shown in the inset of Fig. 9 (*b*). It is quite different from that of $-\Delta S_M^{max}$ which decreases gradually as $x$ increases (Fig. 8). This difference stems from the competing contributions to the RCP from the decreasing $-\Delta S_M^{max}$ and increasing $\delta T_{FWHM}$ as a result of increasing $x$. In addition, it suggests that only the magnetic entropy change is insufficient to justify the potentiality of the practical application for a magnetic refrigerant.

Figure 10 displays a comparison of RCP between $Ga_{1-x}CMn_{3+x}$ ($\Delta H$ = 45 kOe) and other potential candidates for magnetic refrigeration, such as Gd, $Gd_5Si_2Ge_2$, MnAs, $MnFe_{0.35}As_{0.65}$, and $Ni_{0.526}Mn_{0.231}Ga_{0.243}$ ($\Delta H$ = 50 kOe).[23] The RCP of $Ga_{1-x}CMn_{3+x}$ is larger than that of $Ni_{0.526}Mn_{0.231}Ga_{0.243}$ and close to that of $MnFe_{0.35}As_{0.65}$. The largest RCP value of $Ga_{1-x}CMn_{3+x}$ (2.1 J/cm$^3$, corresponding to $x$=0.07 with $T_C$ = 296.5 K) is about 40% of that of Gd. Taking into account the fact that its magnetic moment is much less than that of Gd, the observed RCP in $Ga_{1-x}CMn_{3+x}$ is considerably large. As discussed above, the MCE in $Ga_{1-x}CMn_{3+x}$ is reversible due to its second-order character of the phase transition. Furthermore, the raw materials of $Ga_{1-x}CMn_{3+x}$ are innocuous and relatively low-cost. Therefore, $Ga_{1-x}CMn_{3+x}$, especially with $x$ = 0.07, can become alterative candidates for the magnetic refrigerant in household refrigerators or air conditioning around room temperature.

## 4. Conclusions

Structural and magnetic properties of antiperovskite-structured $Ga_{1-x}CMn_{3+x}$ ($x$=0, 0.06, 0.07, and 0.08) have been investigated experimentally. We find that the Curie temperature $T_c$ and the saturated magnetization increase monotonously with increasing $x$ because the reduction of Mn-Mn atomic distance strengthens the exchange interaction between the neighboring Mn atoms. We have also shown that the reversible MCE occurs near $T_c$ and $-\Delta S_M^{max}$ decreases as $x$ increases. Regardless of the small values of $-\Delta S_M^{max}$, the RCP of $Ga_{1-x}CMn_{3+x}$ are comparable to those of contemporary magnetic refrigerant materials. The considerably large RCP without hysteresis loss, as well as low-cost and innoxious raw materials makes $Ga_{1-x}CMn_{3+x}$ system a promising candidate for the room temperature magnetic refrigerant. In addition, our results show that the magnetic field dependence of $-\Delta S_M^{max}$ can be well understood in terms of the mean field theory.

**Acknowledgments**

This work was supported by the National Key Basic Research under contract No. 2007CB925002, and the National Nature Science Foundation of China under contract No.50701042, No.10774146 and Director's Fund of Hefei Institutes of Physical Science, Chinese Academy of Sciences.

**Figure captions:**

**FIG. 1.** Powder XRD patterns at room temperature for $Ga_{1-x}CMn_{3+x}$ ($x$=0, 0.06, 0.07, and 0.08). Inset: enlargement of XRD pattern around the (220) peak for $Ga_{1-x}CMn_{3+x}$. The asterisk indicates the unreacted graphite.

**FIG. 2.** Temperature dependence of magnetization $M(T)$ under FCC and FCW modes at 100Oe.

**FIG. 3.** Refined lattice parameter $a$ and the Curie temperature $T_C$ for $Ga_{1-x}CMn_{3+x}$.

**FIG. 4.** Isotherm magnetization curves $M(H)$ at 214K for $Ga_{1-x}CMn_{3+x}$. Inset: Saturated magnetic moment as a function of $x$ at 214K.

**FIG. 5.** Isotherm magnetization curves $M(H)$ for $Ga_{1-x}CMn_{3+x}$ around $T_c$ under external magnetic fields up to 45kOe: (*a*) $x$=0; (*b*) $x$=0.06; (*c*) $x$=0.07; (*d*) $x$=0.08. Arrows indicate the increase of temperature with an interval of several Kelvin.

**FIG. 6.** Arrott plots deduced from $M(H)$ curves in Figure 5: (*a*) $x$=0; (*b*) $x$=0.06; (*c*) $x$=0.07; (*d*) $x$=0.08. Arrows indicate the increase of temperature.

**FIG. 7.** Magnetic entropy change $-\Delta S_M$ as a function of temperature under magnetic field changes of $\Delta H$=20kOe (*a*) and 45kOe (*b*) for $Ga_{1-x}CMn_{3+x}$.

**FIG. 8.** Maximum magnetic entropy change $-\Delta S_M^{max}$ at $T_C$ as a function of $H^{2/3}$ for $Ga_{1-x}CMn_{3+x}$

**FIG. 9.**(*a*) Full width at half maximum $\delta T_{FWHM}$ of $-\Delta S_M$ peak for $Ga_{1-x}CMn_{3+x}$ as a function of magnetic field $H$. (*b*) Magnetic field dependence of RCP for all samples. Inset of (*b*), RCP as a function of $x$ at 45 kOe.

**FIG. 10.** A comparison of relative cooling power (RCP) of $Ga_{1-x}CMn_{3+x}$ ($\Delta H$=45 kOe) with those of potential candidates for magnetic refrigerator ($\Delta H$=50 kOe).